# On melting of B$_4$C boron carbide under pressure

Vladimir A. Mukhanov, Petr S. Sokolov and Vladimir L. Solozhenko[*]

*LSPM–CNRS, Université Paris Nord, 93430 Villetaneuse, France*

The pressure dependence of melting temperatures for boron carbide and B$_4$C–carbon eutectic has been studied up to 8 GPa, and it was found that in both cases the melting curves exhibit negative slope (-13±6 K/GPa), that is indicative of higher density of the melt as compared to the solid phase.

Boron carbide B$_4$C [1] is the most studied representative of superhard refractory boron compounds [2,3], to which B$_6$O boron suboxide [4] and recently synthesized B$_{13}$N$_2$ boron subnitride [5] also belong. At ambient pressure the melting temperature of B$_4$C may be evaluated at 2720±50 K [6–8]. However, in the literature there is no data on its melting at high pressures. In the present study the pressure dependence of the melting temperatures of boron carbide and B$_4$C–carbon eutectic at pressures up to 8 GPa was investigated for the first time.

The experiments were conducted in a specially designed high-temperature (up to 3400 K) cell (Fig. 1a) of a toroid-type high-pressure apparatus [9]. The cell was pressure calibrated at room temperature against phase transitions in Bi (2.55 and 7.7 GPa), PbSe (4.2 GPa), and PbTe (5.2 GPa). The temperature calibration under pressure was made using well-established reference points [10]: melting of Si, NaCl, CsCl, Pt, Rh, Al$_2$O$_3$ and Ni–Mn–C ternary eutectic. In the case of Pt, Rh, and Al$_2$O$_3$, the melting points were found in series of quenching experiments from the change of shape and microstructure of the reference material pressed into a cylinder of graphite-like boron nitride, while in the other cases the melting temperature was fixed *in situ* from the electrical resistance jump in the cell (the reference material was in a direct contact with a heater). In the pressure range under study the dependencies of temperature in the center of the cell on the electric power input are linear (Fig. 1b). The line slopes decrease from 2.0 K/W at 2.55 GPa to 1.5 K/W at 7.7 GPa due to an increase of heat losses with increasing pressure. The error of the temperature evaluation in the 2400-2800 K range is ±60 K, and according to the finite element method calculations of temperature fields in the framework of the stationary thermal conductivity problem, the temperature gradients in the sample do not exceed ~15 K/mm in the radial and ~10 K/mm in axial directions.

B$_4$C single crystals (150–200 µm) produced by the interaction of boron oxide (III) with carbon black at 2800 K in the argon atmosphere were used in the experiments. The crystal lattice parameters were *a* = 5.6001(3) Å and *c* = 12.0739 Å, which correspond to the 19.5 at% C composition [1].

The $B_4C$ melting in the 2.5–7.7 GPa pressure range was studied by quenching. The isothermal holding time at a desired pressure was 60–90 s, the cooling rate at the initial stage after switching-off the power was ~300 K/s. No signs of chemical interaction between $B_4C$ and boron nitride capsule were observed over the whole pressure – temperature range under study. Melting was indicated by the formation of a strong monolith sample with a distinct laminar structure, while the samples, which did not attain the melting temperature, remained brittle compacts of relatively large $B_4C$ crystals. The results obtained are shown in Fig. 2a. The lattice parameters of samples quenched from different pressures and temperatures coincide within the limits of experimental error, which is indicative of a congruent melting of boron carbide under pressure.

The $B_4C$ melting in the 4.36–4.55 GPa range was studied also *in situ* by X-ray diffraction with synchrotron radiation using the MAX80 multianvil system at F2.1 beamline of the DORIS III storage ring (HASYLAB-DESY). The experimental details are described elsewhere [11,12], the data obtained are shown in Fig. 2a.

The $B_4C$ melting curve (dashed line in Fig. 2a derived from the results of both quenching and *in situ* experiments) has a negative slope (–13±6 K/GPa) that points to a higher density of the boron carbide melt as compared with the solid phase in the pressure range under study.

In the case of $B_4C$–carbon eutectic the appearance of the liquid phase in the system was fixed *in situ* from a jump of the cell electrical resistance. In this case boron nitride capsules were not used, and the initial mixture of boron carbide and graphite (double excess of carbon as compared with the eutectic composition of ~30 at% C [6,7]) was in the direct contact with the graphite heater. The pressure dependence of the eutectic temperature (dashed line in Fig. 2b) exhibits the same slope (–12±6 K/GPa) as the melting curve of pure $B_4C$.

The authors thank Christian Lathe, Jean-Pierre Michel and Nicolas Fagnon for their help in the preparation of experiments. Financial support from the Agence Nationale de la Recherche (grant ANR-2011-BS08-018-01) is gratefully acknowledged.

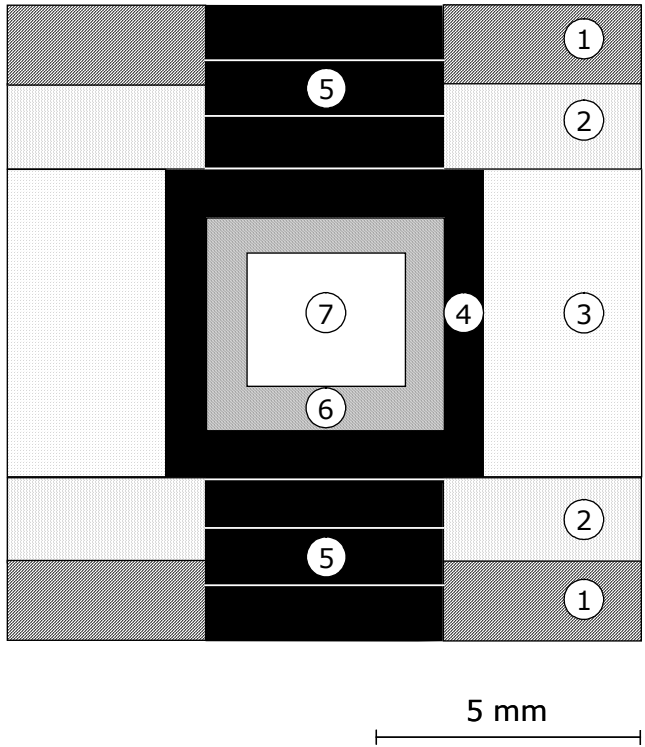 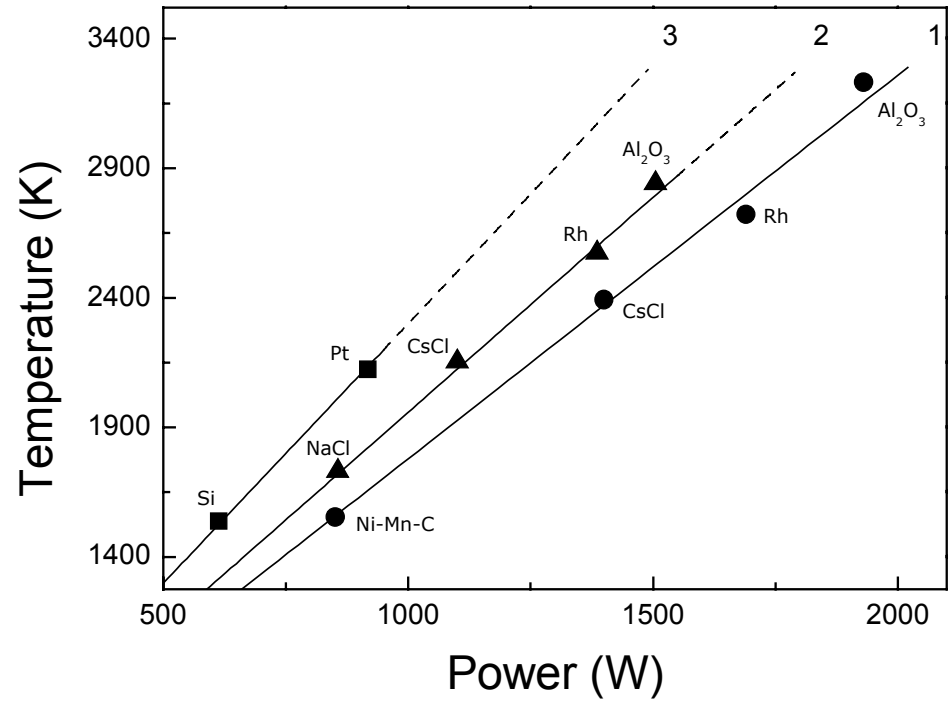

*(a)* *(b)*

**Fig. 1**

*(a)* High-temperature cell of the toroid-type high-pressure apparatus: (*1*) pyrophyllite ring; (*2*) ring of pressed fianite (-100 μm); (*3*) sleeve of pressed fianite (100–200 μm); (*4*) cylindrical graphite heater; (*5*) discs of pressed Ceylon graphite; (*6*) boron nitride capsule; (*7*) sample.

*(b)* Temperature in the center of high-pressure cell *vs* electric power at 7.7 (*1*), 5.2 (*2*) and 2.55 (*3*) GPa.

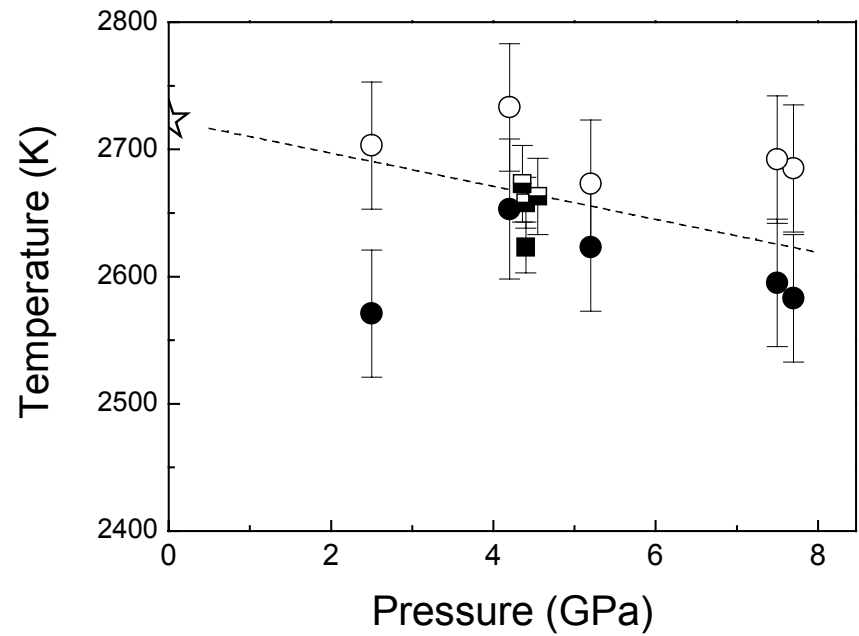 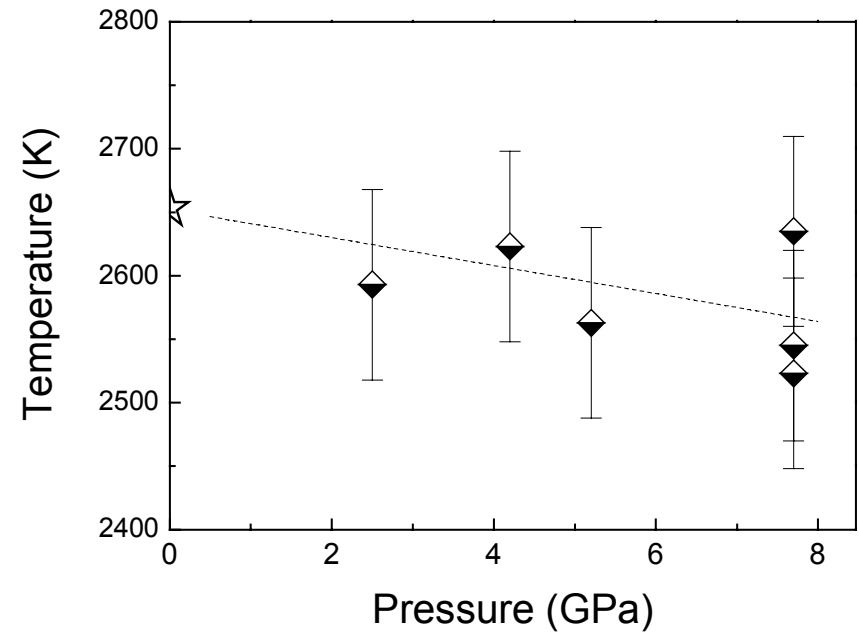

**Fig. 2**

*(a)* Pressure dependence of $B_4C$ melting temperature. The star indicates the melting temperature at ambient pressure (2720±50 K [6–8]); circles show the results of quenching experiments; squares present the results of *in situ* experiments. The open symbols correspond to melting, solid to its absence; half-filled squares indicate the onset of melting registered *in situ* by drastic changes in mutual intensities of reflections in diffraction patterns. Dashed line is the linear approximation of the melting curve defined by least-squares method.

*(b)* Pressure dependence of the $B_4C$–graphite eutectic temperature (dashed line is the linear approximation of the melting curve defined by least-squares method). The star indicates the eutectic temperature at ambient pressure (2650±50 K [7,8]); half-filled diamonds show the results of *in situ* experiments.